\documentclass[12pt]{article}
\usepackage{amsfonts}
\usepackage{mathtools}
\usepackage{slashed}
\usepackage{amsmath,amssymb}
\usepackage{graphicx,color}

\newcommand{\be}{\begin{equation}}
\newcommand{\ee}{\end{equation}}
\newcommand{\bea}{\begin{eqnarray}}
\newcommand{\eea}{\end{eqnarray}}

\begin{document}
%\hfill AEI-2014-012\\

\centerline{\bf \Large Some insights in the structure of correlation functions}
\vspace{3mm}
\centerline{\bf \Large in Liouville and Toda field theories}

\vspace{5mm}
\centerline{\bf }
\vspace{5mm}
\centerline{\bf Parikshit Dutta}
\centerline{pdutta@iiserb.ac.in}
\vspace{5mm}
\begin{center}
{\it Indian Institute of Science Education and Research (Bhopal),\\Indore By-pass Road
Bhauri,
District : Bhopal – 462066,\\
Madhya Pradesh - India.
}
\end{center}
\vspace{35mm}

\begin{abstract}
\noindent
We discuss some aspects of Liouville field theory, starting from operator equation of motion in presence of two screening charges and re-derive the dual zero mode Schwinger Dyson equations for the two screening charges from the path integral. Using functional methods we show the familiar pole structure of Liouville correlation function using the partition function. Next we discuss a generalized structure of the correlation functions obtained from the zero mode functional equations. From this structure we infer the use of the Barnes double Gamma functions to construct a part of the denominator of the correlators and also use Weyl symmetry of the theory to deduce more information about the rest. We similarly extend these arguments in the case of Toda field theories where we make a general statement about the denominator of the three point function and Sine-Liouvile field theory where we only obtain an infinite product structure.

\end{abstract}
%\maketitle

%%%%%%%%%%%%%%%%%%%%%%%%%%%%%%%%%%%%%%%%%%%%%%%%%%
\newpage

\section*{Introduction}
Liouville Field Theory is a very important 2 dimensional field theory and has been studied extensively mostly after it came up in the quantization of non-critical strings by Polyakov \cite{poly}. Although many aspects of this theory is well understood, there still remains open questions in this theory. One of these open questions is the duality that is exhibited by the correlation functions of this theory, namely the $b\to\frac{1}{b}$ duality. Due to this duality there arises reflection symmetry in the spectrum of Liouville theory. There has been many attempts to understand this duality from the path integral perspective \cite{pawloski}. In \cite{pd} we attempted to understand this from the path integral point of view and proposed that renormalizing the bare Schwinger Dyson equation produces two functional equations which the correlation functions satisfy, which gives rise to the duality. This was also understood from the context of screening charges. 

In this article we go a bit deeper. Starting from the operator equation of motion for Liouville field theory in presence of two screening charges \cite{schnittger}, we derive the path integral and the associated zero mode Schwinger Dyson equation which were proposed in \cite{pd}. Then the pole structure of the path integral (correlation functions) is shown to be what one obtains by demanding duality. Next using the zero mode Schwinger Dyson equations we obtain an expression which has information about a part of the correlation function namely the denominator. This part is essentially seen to be an infinite product, having similar structure to the Barnes Double Gamma function suggesting its use in the construction of the denominator

Using similar techniques to Liouville Field Theory we write down the zero mode Schwinger Dyson equations for Toda Field Theory and Sine-Liouville Field Theory. Then an expression is obtained, analogous to Liouville theory i.e. an infinite product. From this structure we get some insights about the structure of the correlation functions in these models. 

\section*{Liouville Equation of motion with two Screening Charges}
In \cite{schnittger}, it is shown that the quantum operator equation of motion (on the euclidean cylinder) satisfied by the Liouville field in the presence of two Screening charge operators is given by:
\begin{equation}\label{eqm}
\Box\tilde{\Phi}(x)=\mu 2b e^{2b\phi_{+}(x)}+\tilde{\mu}\frac{2}{b}e^{\frac{2}{b}\phi_{-}(x)}
\end{equation}
Where the field $\tilde{\Phi}(x)=\phi_{+}(x)+\phi_{-}(x)+v_{1}(x)+\bar{v}_{1}(x)$, is the Quantum Liouville field and $v_{1}(x)$ and $\bar{v}_{1}(x)$ are holomorphic and antiholomorphic free fields respectively, while $\phi_{-}(x)$ and $\phi_{+}(x)$ satisfy the equations of motion with single exponent i.e. in presence of a single screening charge and is parametrized using the free fields ($v_{1},\,\bar{v}_{1}$). This is an interesting operator equation and one can try to compare this to our construction \cite{pd}. The motivation is then to write down the path integral for this operator equation. To do this we invoke the Schwinger action principle, which sates that the path integral is the solution of the expectation value of the operator equation of motion. The expectation value of this operator equation in presence of a source term is:
\begin{equation}
\bigg\langle\Box\tilde{\Phi}(x)-\mu 2b e^{2b\phi_{+}(x)}-\tilde{\mu}\frac{2}{b}e^{\frac{2}{b}\phi_{-}(x)}+\tilde{J}\bigg\rangle=0
\end{equation}
Now, we would like to write the solution to the expectation value of the equation of motion above (which is the Schwinger Dyson equation in some sense). Firstly let us define this equation on the sphere :
\begin{equation}\label{mastereq}
\bigg\langle\frac{1}{2\pi}\Delta\tilde{\Phi}(x)-\frac{Q}{4\pi}R(x)-\mu 2b e^{2b\phi_{+}(x)}-\tilde{\mu}\frac{2}{b}e^{\frac{2}{b}\phi_{-}(x)}+\tilde{J}\bigg\rangle=0
\end{equation}
Where now the field $\tilde{\Phi}(x)=\phi_{+}(x)+\phi_{-}(x)-\phi_{0}(x)$ , where $\phi_{0}$ is a free field (equivalent to $v_{1}+\bar{v}_{1}$) and $\Delta$ is the Laplace-Beltrami operator.  
\subsection*{Path Integral}

The solution to the expectation value of the operator equation of motion, can be written as the path integral \cite{fujikawa}:
\begin{equation}\label{pi}
Z[J_{1},J_{2},J_{3}]=\int [D\phi_{+}][D\phi_{-}][D\phi_{0}]e^{\int d^{2}x\sqrt{g}J_{1}\phi_{+}+\int d^{2}x\sqrt{g}J_{2}\phi_{-}+\int d^{2}x \sqrt{g} J_{3}\phi_{0}-S[\phi_{+},\phi_{-},\phi_{0}]}
\end{equation}
Where the action is defined as:
\begin{equation}
S[\phi_{+},\phi_{-},\phi_{0}]=\frac{1}{4\pi}\int d^{2}x\sqrt g \left(g^{kl}\partial_k\tilde{\Phi}\,\,\partial_l\tilde{\Phi}+Q\,R\tilde{\Phi}+4\pi\mu e^{2b\phi_{+}}+4\pi\tilde{\mu}e^{\frac{2}{b}\phi_{-}}\right)
\end{equation}
Now the Liouville correlator corresponds to the condition, $J_{1}=J_{2}=-J_{3}=\tilde{J}$ or $(J_{1}+J_{2}+J_{3})$. Since the free field $\phi_{0}$ is used to parametrize the fields $\phi_{+}$ and $\phi_{-}$ in the operator approach, the functional integral is non-trivial. To perform the integration over the fields $\phi_{+},\,\phi_{-},\,\phi_{0}$, we implicitly enforce the constraint over the sources, i.e $J_{1}=J_{2}=-J_{3}$, and then the fields can be treated as independent variables. As any functional of the fields can be replaced by functional of the derivative with respect to the sources, putting the sources, $J_{1}=J_{2}=-J_{3}$ in the partition function (while performing the computation) allows to keep the non-trivial relation between the fields.  Also it is to be noted that, for the path integral for the Liouville correlator, shift in any one of the sources (e.g. for the choice of  $J_{1}+\alpha,J_{2},-J_{3}$) must be compensated by shift in the other two sources  (i.e. should be same as, $J_{1}+\alpha,J_{2}+\alpha,-J_{3}-\alpha$ to satisfy the condition stated above, $\alpha$ being any arbitrary function) as this solves the same equation of motion (\ref{mastereq}). In short the fields can be treated as independent (the measure translation invariant) if all the sources are simultaneously shifted.

 The Liouville correlation functions then correspond to the case,
\begin{equation}\label{de1}
\tilde{J}=J_{1}(x)=J_{2}(x)=-J_{3}(x)=\sum_i^n \frac{2\alpha_{i}}{\sqrt{ g(x)}}\,\delta^{2}(x-x_{i})
\end{equation}
Let us see how we get back eq (\ref{mastereq}) from this. Performing translation invariance of the measure with respect to each of the fields $\phi_{+}$, $\phi_{-}$,$\phi_{0}$, one obtains the following three equations:
\begin{eqnarray}
&& \!\!\!\!\!\! \!\!\!\!\!\! \!\!\!\!\!\!\!\!\!\!\!\! \!\!\!\!\!\!\!\!\!\!\!\
\bigg\langle\frac{1}{2\pi}\Delta\tilde{\Phi}(x)-\frac{Q}{4\pi}R(x)-\mu 2b e^{2b\phi_{+}(x)}+J_{1}(x)\bigg\rangle=0\\
&& \!\!\!\!\!\! \!\!\!\!\!\! \!\!\!\!\!\!\!\!\!\!\!\! \!\!\!\!\!\!\!\!\!\!\!\
\bigg\langle\frac{1}{2\pi}\Delta\tilde{\Phi}(x)-\frac{Q}{4\pi}R(x)-\tilde{\mu}\frac{2}{b}e^{\frac{2}{b}\phi_{-}(x)}+J_{2}(x)\bigg\rangle=0\\
&& \!\!\!\!\!\! \!\!\!\!\!\! \!\!\!\!\!\!\!\!\!\!\!\! \!\!\!\!\!\!\!\!\!\!\!\
-\bigg\langle\frac{1}{2\pi}\Delta\tilde{\Phi}(x)+\frac{Q}{4\pi}R(x)+J_{3}(x)\bigg\rangle=0
\end{eqnarray}
Now taking the sum of the above equations, and for the condition $J_{1}=J_{2}=-J_{3}=\tilde{J}$, we get the equation for the Liouville correlator i.e. eq(\ref{mastereq}). We now do a zero mode splitting as earlier \cite{pd} of the two fields, 
\begin{equation}
\phi_{+}(x)=\phi_{+}^{0}+\tilde{\phi}_{+}(x)\quad {\rm and}\quad\phi_{-}(x)=\phi_{-}^{0}+\tilde{\phi}_{-}(x)
\end{equation}
$\phi_{+}^{0},\phi_{-}^{0}$ are the respective zero modes, and so constant on the sphere. Performing translation invariance of the measure with respect to the two zero modes one obtains the following:
\begin{eqnarray}\label{sc1/b}
&& \!\!\!\!\!\! \!\!\!\!\!\! \!\!\!\!\!\!\!\!\!\!\!\! \!\!\!\!\!\!\!\!\!\!\!\
\int d^{2}x\sqrt{g}(J_{2}(x)-\frac{Q}{4\pi}R(x))Z[J_{1},J_{2},J_{3}]=\tilde{\mu}\frac{1}{b}\int d^{2}x \sqrt{g}Z[J_{1},J_{2,x,1/b},J_{3}]\\
&& \!\!\!\!\!\! \!\!\!\!\!\! \!\!\!\!\!\!\!\!\!\!\!\! \!\!\!\!\!\!\!\!\!\!\!\
\int d^{2}x\sqrt{g}(J_{1}(x)-\frac{Q}{4\pi}R(x))Z[J_{1},J_{2},J_{3}]=\mu b\int d^{2}x \sqrt{g}Z[J_{1,x,b},J_{2},J_{3}]
\end{eqnarray}
 Where,
 \begin{eqnarray}
&& \!\!\!\!\!\! \!\!\!\!\!\! \!\!\!\!\!\!\!\!\!\!\!\! \!\!\!\!\!\!\!\!\!\!\!\
J_{2,x,1/b}(y)=J_{2}(y)+\frac{2}{b}\frac{1}{\sqrt{g(x)}}\delta^{2}(y-x)\\
&& \!\!\!\!\!\! \!\!\!\!\!\! \!\!\!\!\!\!\!\!\!\!\!\! \!\!\!\!\!\!\!\!\!\!\!\
J_{1,x,b}(y)=J_{1}(y)+\frac{2b}{\sqrt{g(x)}}\delta^{2}(y-x)
\end{eqnarray}
But as we pointed out earlier, the Liouville correlator satisfies the equation (\ref{mastereq}), and this equation is invariant under simultaneous redefinitions of either of $J_{1}$ or $J_{2}$ with respect to $J_{3}$. Moreover as pointed out earlier we must satisfy the constraint $J_{1}=J_{2}=-J_{3}=\tilde{J}$. And so we have the the equation for $J_{1}=J_{2}=-J_{3}$, (we will also show this by performing the functional integral explicitly in the next section):
\begin{eqnarray}
&& \!\!\!\!\!\! \!\!\!\!\!\! \!\!\!\!\!\!\!\!\!\!\!\! \!\!\!\!\!\!\!\!\!\!\!\
\int d^{2}x\sqrt{g}(J_{2}(x)-\frac{Q}{4\pi}R(x))Z[J_{1},J_{2},J_{3}]=\tilde{\mu}\frac{1}{b}\int d^{2}x \sqrt{g}Z[J_{1,x,1/b},J_{2,x,1/b},J_{3,x,1/b}]\\
&& \!\!\!\!\!\! \!\!\!\!\!\! \!\!\!\!\!\!\!\!\!\!\!\! \!\!\!\!\!\!\!\!\!\!\!\
\int d^{2}x\sqrt{g}(J_{1}(x)-\frac{Q}{4\pi}R(x))Z[J_{1},J_{2},J_{3}]=\mu b\int d^{2}x \sqrt{g}Z[J_{1,x,b},J_{2,x,b},J_{3,x,b}]
\end{eqnarray}
Thus the equations become, for the choice of source (\ref{de1}):
\begin{eqnarray}\label{receq1}
&& \!\!\!\!\!\! \!\!\!\!\!\! \!\!\!\!\!\!\!\!\!\!\!\! \!\!\!\!\!\!\!\!\!\!\!\
[\tilde{\alpha}-Q]\langle\prod_{i=1}^{n}e^{2\alpha_{i}\tilde{\Phi}(x_{i})}\rangle=\mu b \int d^{2}x \sqrt{ g(x)}\langle \prod_{i=1}^{n}e^{2\alpha_{i}\tilde{\Phi}(x_{i})}e^{2b\tilde{\Phi}(x)}\rangle\\
&& \!\!\!\!\!\! \!\!\!\!\!\! \!\!\!\!\!\!\!\!\!\!\!\! \!\!\!\!\!\!\!\!\!\!\!\  \label{receq2}
[\tilde{\alpha}-Q]\langle\prod_{i=1}^{n}e^{2\alpha_{i}\tilde{\Phi}(x_{i})}\rangle=\tilde{\mu} \frac{1}{b} \int d^{2}x \sqrt{g(x)}\langle \prod_{i=1}^{n}e^{2\alpha_{i}\tilde{\Phi}(x_{i})}e^{\frac{2}{b}\tilde{\Phi}(x)}\rangle
\end{eqnarray}
Which are the two separate screening charge equations which was argued in \cite{pd} as the source of the duality for the quantum Liouville theory. It is to be noted that in \cite{pd} the two functional equations were obtained from renormalizing the bare Schwinger Dyson equation while here it is taken into account from the beginning that due to renormalization there exist two screening charges in the theory and hence the equation of motion from which the path integral is constructed has explicitly two screening charges in it. This also produces the same functional equations as earlier. Also one must note that the L.H.S. is same for eq. (\ref{receq1}) and (\ref{receq2}), meaning the R.H.S. is same as well. This means that the insertion of the two screening charges along with interchange of $\mu$ and $\tilde{\mu}$, is equivalent to each other, inside the expectation value of primary fields. Since this holds for any number of primary field  insertions and arbitrary $\alpha$s it is clear that the two screening charge operators are related to each other upto a constant (which is the reflection symmetry of the Weyl group as we will also discuss later).

\subsection*{Pole structure from path integral}

 One can go further and try to compute the path integral more explicitly. Let us relax the constraint on the sources for the moment. From the definition of the path integral, we can do the following algebra, using the fact that any polynomial (functional) of the field can be written as a polynomial of the derivative with respect to the external source $J$:
\begin{eqnarray}
&& \!\!\!\!\!\! \!\!\!\!\!\! \!\!\!\!\!\!\!\!\!\!\!\! \!\!\!\!\!\!\!\!\!\!\!\
Z[J_{1},J_{2},J_{3}]=e^{-\int d^{2}x\sqrt{g}\mu e^{2b\frac{\delta}{\delta J_{1}(x)}}}e^{-\int d^{2}x\sqrt{g}\tilde{\mu}e^{\frac{2}{b}\frac{\delta}{\delta J_{2}(x)}}}\times\notag\\
&& \!\!\!\!\!\! \!\!\!\!\!\! \!\!\!\!\!\!\!\!\!\!\!\! \!\!\!\!\!\!\!\!\!\!\!\
\int [D\phi_{+}][D\phi_{-}][D\phi_{0}]e^{-\frac{1}{4\pi}\int d^{2}x\sqrt g (g^{kl}\partial_k\tilde{\Phi}\,\,\partial_l\tilde{\Phi}+Q\,R\tilde{\Phi})+\int d^{2}x\sqrt{g}(J_{1}\phi_{+}+J_{2}\phi_{-}+J_{3}\phi_{0})}\notag
\end{eqnarray}
Where, $\tilde{\Phi}(x)=\phi_{+}(x)+\phi_{-}-\phi_{0}(x)$. Now let us look at the term in exponent inside the path integral in detail: 
\begin{eqnarray}
&& \!\!\!\!\!\! \!\!\!\!\!\! \!\!\!\!\!\!\!\!\!\!\!\! \!\!\!\!\!\!\!\!\!\!\!\
-\frac{1}{4\pi}\int d^{2}x\sqrt g (g^{kl}\partial_k\tilde{\Phi}\,\,\partial_l\tilde{\Phi}+Q\,R\tilde{\Phi})+(J_{1}(x)\phi_{+}(x)+J_{2}(x)\phi_{-}(x)+J_{3}(x)\phi_{0}(x))\notag\\
&& \!\!\!\!\!\! \!\!\!\!\!\! \!\!\!\!\!\!\!\!\!\!\!\! \!\!\!\!\!\!\!\!\!\!\!\
=-\frac{1}{4\pi}\int d^{2}x\sqrt g (g^{kl}\partial_k\tilde{\Phi}\,\,\partial_l\tilde{\Phi})+(J_{1}-\frac{Q}{4\pi}R(x))\phi_{+}+(J_{2}-\frac{Q}{4\pi}R(x))\phi_{-}+(J_{3}+\frac{Q}{4\pi}R(x))\phi_{0}\notag\\
\end{eqnarray}
Changing the variables of integration in the path integral as follows:
\begin{equation}
\phi_{+}\to\tilde{\Phi}=\phi_{+}+\phi_{-}-\phi_{0}\quad\quad\phi_{-}\to\phi_{-}\quad\quad\phi_{0}\to\phi_{0}
\end{equation}
(the Jacobian of the above transformation is a constant since it is a linear transformation). The path integral can then be written as follows:
\begin{eqnarray}
&& \!\!\!\!\!\! \!\!\!\!\!\! \!\!\!\!\!\!\!\!\!\!\!\! \!\!\!\!\!\!\!\!\!\!\!\
Z[J_{1},J_{2},J_{3}]=e^{-\int d^{2}x\sqrt{g}\mu e^{2b\frac{\delta}{\delta J_{1}(x)}}}e^{-\int d^{2}x\sqrt{g}\tilde{\mu}e^{\frac{2}{b}\frac{\delta}{\delta J_{2}(x)}}}\times\notag\\
&& \!\!\!\!\!\! \!\!\!\!\!\! \!\!\!\!\!\!\!\!\!\!\!\! \!\!\!\!\!\!\!\!\!\!\!\
\int [D\tilde{\Phi}][D\phi_{-}][D\phi_{0}]e^{-\frac{1}{4\pi}\int d^{2}x\sqrt g (g^{kl}\partial_k\tilde{\Phi}\,\,\partial_l\tilde{\Phi})+\int d^{2}x\sqrt{g}((J_{1}-\frac{Q}{4\pi}R(x))\tilde{\Phi}+(J_{2}-J_{1})\phi_{-}+(J_{1}+J_{3})\phi_{0})}\notag
\end{eqnarray}

From the above calculations it is clear that the integral over $\phi_{-}$, and $\phi_{0}$ become Lagrange multipliers and so we obtain the constraints after these two integrations: $J_{1}=J_{2},J_{1}=-J_{3}$. Analogously if one now puts the constraint $J_{1}=J_{2}=-J_{3}$ back in the functional integral, the integral over $\phi_{-}$ and $\phi_{0}$ decouples completely from the rest and hence are no longer important in the partition function. Thus the path integral can then be written as: 

\begin{eqnarray}\label{part1}
&& \!\!\!\!\!\! \!\!\!\!\!\! \!\!\!\!\!\!\!\!\!\!\!\! \!\!\!\!\!\!\!\!\!\!\!\
Z[J_{1},J_{1},-J_{1}]=e^{-\int d^{2}x\sqrt{g}\mu e^{2b\frac{\delta}{\delta J_{1}(x)}}}e^{-\int d^{2}x\sqrt{g}\tilde{\mu}e^{\frac{2}{b}\frac{\delta}{\delta J_{1}(x)}}}\times\notag\\
&& \!\!\!\!\!\! \!\!\!\!\!\! \!\!\!\!\!\!\!\!\!\!\!\! \!\!\!\!\!\!\!\!\!\!\!\
\int [D\tilde{\Phi}]e^{-\frac{1}{4\pi}\int d^{2}x\sqrt g (g^{kl}\partial_k\tilde{\Phi}\,\,\partial_l\tilde{\Phi})+\int d^{2}x\sqrt{g}((J_{1}-\frac{Q}{4\pi}R(x))\tilde{\Phi}}\notag\\
&& \!\!\!\!\!\! \!\!\!\!\!\! \!\!\!\!\!\!\!\!\!\!\!\! \!\!\!\!\!\!\!\!\!\!\!\
=e^{-\int d^{2}x\sqrt{g}\mu e^{2b\frac{\delta}{\delta J_{1}(x)}}}e^{-\int d^{2}x\sqrt{g}\tilde{\mu}e^{\frac{2}{b}\frac{\delta}{\delta J_{1}(x)}}}Z_{0}[J_{1}]_{Q}\notag\\
\end{eqnarray}
Where $Z_{0}[J_{1}]_{Q}$, is a free field Coloumb Gas integral with background charge $Q$, explicitly written for delta function sources as:
\begin{eqnarray}
&& \!\!\!\!\!\! \!\!\!\!\!\! \!\!\!\!\!\!\!\!\!\!\!\! \!\!\!\!\!\!\!\!\!\!\!\
Z_{0}[J_{1}]_{Q}=\int [D\tilde{\Phi}]e^{-\frac{1}{4\pi}\int d^{2}x\sqrt g (g^{kl}\partial_k\tilde{\Phi}\,\,\partial_l\tilde{\Phi})+\int d^{2}x\sqrt{g}((J_{1}-\frac{Q}{4\pi}R(x))\tilde{\Phi}}\notag\\
&& \!\!\!\!\!\! \!\!\!\!\!\! \!\!\!\!\!\!\!\!\!\!\!\! \!\!\!\!\!\!\!\!\!\!\!\
=e^{\int\int d^{2}x\sqrt{g(x)}\,d^{2}y\sqrt{g(y)}(J_{1}(x)-\frac{Q}{4\pi}R(x))(\frac{1}{4\pi}\Delta)^{-1}(J_{1}(y)-\frac{Q}{4\pi}R(y))}\notag\\
&& \!\!\!\!\!\! \!\!\!\!\!\! \!\!\!\!\!\!\!\!\!\!\!\! \!\!\!\!\!\!\!\!\!\!\!\
=\langle \prod_{i=1}^{n}e^{2\alpha_{i}\tilde{\Phi}(x_{i})}\rangle_{C.G.}\notag
\end{eqnarray}
In the above, the expressions are implicitly understood to be regularized by removing self contractions of the exponentials and their cross terms. If one now expands eq (\ref{part1}) in a Taylor series then one obtains the following:
\begin{eqnarray}
Z[J_{1},J_{1},-J_{1}]&&=\sum_{k=0}^{\infty}\frac{1}{k!}\bigg(-\int d^{2}x\sqrt{g}\mu e^{2b\frac{\delta}{\delta J_{1}(x)}}\bigg)^{k} \sum_{l=0}^{\infty}\frac{1}{l!}\bigg(-\int d^{2}x\sqrt{g}\tilde{\mu}e^{\frac{2}{b}\frac{\delta}{\delta J_{1}(x)}}\bigg)^{l}Z_{0}[J_{1}]_{Q}\notag\\
&&=\sum_{k=0}^{\infty}\frac{(-\mu)^{k}}{k!}\sum_{l=0}^{\infty}\frac{(-\tilde{\mu})^{l}}{l!}\langle \prod_{i}^{n}e^{2\alpha_{i}\tilde{\Phi}(x_{i})}{\cal{Q}}_{+}^{k}{\cal{Q}}_{-}^{l}\rangle_{C.G.}\notag\\
\end{eqnarray}
Where,
\begin{equation}
{\cal{Q}}_{+}=\int d^{2}x\sqrt{g(x)} e^{2b\tilde{\Phi}(x)}\quad{\cal{Q}}_{-}=\int d^{2}x \sqrt{g(x)} e^{\frac{2}{b}\tilde{\Phi}(x)}
\end{equation}
are the two screening charges. Now we can explicitly see from the above equation that each of the terms in the expansion correspond to the residues of the Liouville correlation function at the points, $\tilde{\alpha}-Q=-kb-\frac{l}{b}$ for $k,l$ integers, $\geq 0$.

We note that the pole expansion is identical to the one obtained in \cite{pawloski}, for the two exponential potential. In their case they had introduced the two exponential potential by hand, where as in our case we showed that renormalizing the bare Schwinger Dyson equation produces two screening charge equations for the correlation functions \cite{pd} . We also show here that starting from the operator equation of motion for two screening charges, we obtain the same functional equation as in \cite{pd}, and thus the two approaches are equivalent. From the pole expansion one can perform the same trick as in \cite{pawloski} to use the Sommerfeld-Watson transform and write the pole expansion as a contour integral. From there one can clearly see that the analytic continuation of the three point function form the residues leads to the DOZZ formula, with a possible extra periodic function $g(\tilde{\alpha}-Q)$ such that $g(-mb-n/b)=1$. for irrational $b$ the periods are incommensurate with each other and the formula is unique. Also the other point is that the sum of equations (\ref{receq1}),(\ref{receq2}) is equivalent to the equation employed by \cite{pawloski} to prove the uniqueness of the three point function. We shall not try to reproduce their calculations here. Once the three point function is fixed, then one can see that there is a reflection symmetry $\alpha\to Q-\alpha$ of the function. From the two point function analysis, it was argued that due to this reflection symmetry, there exists only one state for each conformal dimension $\alpha(Q-\alpha)$. This is manifestation of the Weyl group symmetry of Sl(2) Toda field theory which is Liouville theory.

\subsection*{Infinite product and the Double Gamma function}
The recursive relations for the correlation functions (\ref{receq1}) and (\ref{receq2}) can be used to gain more information. Using the equation (\ref{receq1}) we can write the following analogous to \cite{pd},
\begin{eqnarray}\label{it1}
&& \!\!\!\!\!\! \!\!\!\!\!\! \!\!\!\!\!\!\!\!\!\!\!\! \!\!\!\!\!\!\!\!\!\!\!\!
\prod_{j=0}^{m}(\tilde{\alpha}+j b-Q)\,\,\Big\langle\prod_{i=1}^{n}e^{2\alpha_{i}\phi(x_{i})}\Big\rangle\notag\\
&& \!\!\!\!\!\! \!\!\!\!\!\! \!\!\!\!\!\!\!\!\!\!\!\! \!\!\!\!\!\!\!\!\!\!\!\!
\quad \quad \quad \quad \quad\quad\quad =(b\mu )^{m+1}\idotsint\Big\langle\prod_{i=1}^{n}e^{2\alpha_{i}\phi(x_{i})}\prod_{j=0}^{m}e^{2b\phi(y_{j})}\Big\rangle d^{2}y_{j}\sqrt{g(y_{j})}~.
\end{eqnarray}
 We use $\phi(x)$ as the Liouville field here for simplicity. Now if one has an extra insertion of $\alpha=1/b$ then we have the following:
\begin{eqnarray}
&& \!\!\!\!\!\! \!\!\!\!\!\! \!\!\!\!\!\!\!\!\!\!\!\! \!\!\!\!\!\!\!\!\!\!\!\!
\prod_{j=0}^{m}(\tilde{\alpha}+\frac{1}{b}+j b-Q)\,\,\Big\langle\prod_{i=1}^{n}e^{2\alpha_{i}\phi(x_{i})}e^{\frac{2}{b}\phi(x)}\Big\rangle\notag\\
&& \!\!\!\!\!\! \!\!\!\!\!\! \!\!\!\!\!\!\!\!\!\!\!\! \!\!\!\!\!\!\!\!\!\!\!\!
\quad \quad \quad \quad \quad\quad\quad =(b\mu )^{m+1}\idotsint\Big\langle\prod_{i=1}^{n}e^{2\alpha_{i}\phi(x_{i})}e^{\frac{2}{b}\phi(x)}\prod_{j=0}^{m}e^{2b\phi(y_{j})}\Big\rangle d^{2}y_{j}\sqrt{g(y_{j})}~.
\end{eqnarray}
Performing integration over $x$, we have:
\begin{eqnarray}
&& \!\!\!\!\!\! \!\!\!\!\!\! \!\!\!\!\!\!\!\!\!\!\!\! \!\!\!\!\!\!\!\!\!\!\!\!
\prod_{j=0}^{m}(\tilde{\alpha}+\frac{1}{b}+j b-Q)\,\int\,d^{2}x\sqrt{g(x)}\,\Big\langle\prod_{i=1}^{n}e^{2\alpha_{i}\phi(x_{i})}e^{\frac{2}{b}\phi(x)}\Big\rangle\notag\\
&& \!\!\!\!\!\! \!\!\!\!\!\! \!\!\!\!\!\!\!\!\!\!\!\! \!\!\!\!\!\!\!\!\!\!\!\!
\quad \quad \quad \quad =(b\mu )^{m+1}\int\,d^{2}x\sqrt{g(x)}\idotsint\Big\langle\prod_{i=1}^{n}e^{2\alpha_{i}\phi(x_{i})}e^{\frac{2}{b}\phi(x)}\prod_{j=0}^{m}e^{2b\phi(y_{j})}\Big\rangle d^{2}y_{j}\sqrt{g(y_{j})}~.
\end{eqnarray}

The L.H.S. of above can be rewritten using (\ref{receq2}), and hence we obtain:
\begin{eqnarray}
&& \!\!\!\!\!\! \!\!\!\!\!\! \!\!\!\!\!\!\!\!\!\!\!\! \!\!\!\!\!\!\!\!\!\!\!\!
\prod_{j=0}^{m}\prod_{l=0}^{1}(\tilde{\alpha}+l\frac{1}{b}+j b-Q)\,\,\Big\langle\prod_{i=1}^{n}e^{2\alpha_{i}\phi(x_{i})}\Big\rangle\notag\\
&& \!\!\!\!\!\! \!\!\!\!\!\! \!\!\!\!\!\!\!\!\!\!\!\! \!\!\!\!\!\!\!\!\!\!\!\!
\quad \quad \quad \quad =(b\mu )^{m+1}(\frac{\tilde{\mu}}{b})\int\,d^{2}x\sqrt{g(x)}\idotsint\Big\langle\prod_{i=1}^{n}e^{2\alpha_{i}\phi(x_{i})}e^{\frac{2}{b}\phi(x)}\prod_{j=0}^{m}e^{2b\phi(y_{j})}\Big\rangle d^{2}y_{j}\sqrt{g(y_{j})}~.
\end{eqnarray}

This is again an iterative equation and one can perform this infinite times. Thus after large number of iterations:
\begin{eqnarray}
&& \!\!\!\!\!\! \!\!\!\!\!\! \!\!\!\!\!\!\!\!\!\!\!\! \!\!\!\!\!\!\!\!\!\!\!\!
\lim_{m,p\to\infty}\prod_{j=0}^{m}\prod_{l=0}^{p}(\tilde{\alpha}+l\frac{1}{b}+j b-Q)\,\,\Big\langle\prod_{i=1}^{n}e^{2\alpha_{i}\phi(x_{i})}\Big\rangle\notag\\
&& \!\!\!\!\!\! \!\!\!\!\!\! \!\!\!\!\!\!\!\!\!\!\!\! \!\!\!\!\!\!\!\!\!\!\!\!
 =\lim_{m,p\to\infty}(b\mu )^{m+1}(\frac{\tilde{\mu}}{b})^{p}\idotsint\Big\langle\prod_{i=1}^{n}e^{2\alpha_{i}\phi(x_{i})}\prod_{l=0}^{p}e^{\frac{2}{b}\phi(x_{l})}\prod_{j=0}^{m}e^{2b\phi(y_{j})}\Big\rangle d^{2}y_{j}\sqrt{g(y_{j})}\int\,d^{2}x_{l}\sqrt{g(x_{l})}~.
\end{eqnarray}
The infinite product appearing on the L.H.S. suggest that the denominator of the correlation function in the L.H.S. must have a similar infinite product for the R.H.S. to be finite at the zeroes of the infinite product, which is true because the R.H.S. reduces to Fateev Dotsenko integrals at these points. Now
\begin{equation}
\prod_{j=0}^{\infty}\prod_{l=0}^{\infty}(\tilde{\alpha}+l\frac{1}{b}+j b-Q)
\end{equation}
is a part of the inverse of the Barnes Double Gamma function $\Gamma_{2}(x;b,\frac{1}{b})$, \cite{barnes}, which is an entire function. This is infact a consequence of Weistrass factorization theorem for entire functions which says that any entire function can be factorized into an infinite product over its zeroes on the complex plane and an additional part (which is entire in itself). It will be interesting to see whether this infinite product can be utilized to write the part of the denominator of the 3-point function in Liouville theory, which is the piece as a function of $(\tilde{\alpha}-Q)$. Since the rest of the part, i.e. the residues can be written using the $\Upsilon$ functions \cite{pawloski} and admits direct analytic continuation, one can expect that deducing the part dependent functionally on $\tilde{\alpha}-Q$, would lead to a direct computation of the 3-point function. This is a problem we would like to look at in the future. Also we note that the zero mode part of the path integral converges only for $\tilde{\alpha}-Q\ge 0$, which is similar to the integral representation of the Gamma function for complex numbers. It would be interesting to see if the zero mode integration can be deformed by choosing an integral representation which converges also for $\tilde{\alpha}-Q\le 0$. 
From the reflection symmetry of the Weyl group of Sl(2), which was discussed earlier, it can be seen that there must exist a dual part to $\Gamma_{2}(\tilde{\alpha}-Q)$. This is given by $\Gamma_{2}(2Q-\tilde{\alpha})$ which would have a part as as infinite product expansion of the form :
\begin{equation}
\prod_{j=0}^{\infty}\prod_{l=0}^{\infty}(2Q-\tilde{\alpha}+l\frac{1}{b}+j b)
\end{equation}
This manifestly has the other poles at $\tilde{\alpha}=2Q+l/b+jb$. for non negative integers $l,j$. It is interesting to note that the inverse of the product of the two Barnes double Gamma functions is the Upsilon function $\Upsilon(x)$ upto a constant. This is analogous to the Euler reflection formula for the simple Gamma function : 
\begin{equation}
\frac{sin(\pi z)}{\pi}=\frac{1}{\Gamma(z)\Gamma(1-z)}
\end{equation}
Thus the replacement of the inverse double Gamma function by the Upsilon function is to choose an entire function (similar to $sin(\pi z)$) which is also invariant under symmetry of the Weyl group of Sl(2). We will later discuss a bit more about the the rest of the double gamma functions in the denominator of the full three point function, which can be understood to be arrising from Weyl group symmetry as well.

\section*{Sl(n) Toda field theories}

Now we will try to discuss about Sl(n) Toda field theories and use analogous concepts to Liouville theory in order to understand it. The action for conformaly invariant Toda field theories is given as follows \cite{Fateev}:
\begin{equation}
A_{TFT}=\int \bigg(\frac{1}{8\pi}g^{\mu\nu}(\partial_{\mu}\varphi,\partial_{\nu}\varphi)+\frac{(Q,\varphi)}{4\pi}R+\mu\sum_{k=1}^{n-1}e^{b(e_{k},\varphi)}\bigg)\sqrt{g}\,d^{2}x
\end{equation}
Where $R$ is the scalar curvature of the background metric and $(\, , \,)$ denote a scalar product with the matrix of scalar products $K_{[ij]}=(e_{i},e_{j})$ (indices in the square bracket mean matrix indices). The background charge is given by:
\begin{equation}
Q=\bigg(b+\frac{1}{b}\bigg)\rho
\end{equation}
and $\rho$ is a Weyl vector (half of the sum of all positive roots). The primary operators of this theory are exponential vertex operators parametrized by a $(n-1)$ component vector parameter $\alpha$ :
\begin{equation}
V_{\alpha}(z,\bar{z})=e^{(\alpha,\varphi(z,\bar{z}))}
\end{equation}
Now the correlation functions are defined using the path integral as follows:
\begin{equation}
\langle V_{\alpha_{1}}(z_{1},\bar{z}_{1})...V_{\alpha_{l}}(z_{l},\bar{z}_{l})\rangle=\int[D\varphi]e^{-A_{TFT}} V_{\alpha_{1}}(z_{1},\bar{z}_{1})...V_{\alpha_{l}}(z_{l},\bar{z}_{l})
\end{equation}
Now each of the component fields $\varphi_{i}$ can be split up into a zero mode $\varphi_{i}^{0}$ and fluctuating part $\tilde{\varphi}_{i}$. Translation invariance of the measure with respect to the zero modes yield the following equations:
\begin{eqnarray}
&& \!\!\!\!\!\! \!\!\!\!\!\! \!\!\!\!\!\!\!\!\!\!\!\! \!\!\!\!\!\!\!\!\!\!\!\!
(\sum_{j=1}^{l}\alpha_{j[m]}-2Q_{[m]})K_{[mi]}\langle V_{\alpha_{1}}(z_{1},\bar{z}_{1})...V_{\alpha_{l}}(z_{l},\bar{z}_{l})\rangle\notag\\
&& \!\!\!\!\!\! \!\!\!\!\!\! \!\!\!\!\!\!\!\!\!\!\!\! \!\!\!\!\!\!\!\!\!\!\!\!
\quad\quad\quad\quad\quad\quad=\mu b \sum_{k=1}^{n-1}e_{k[m]}K_{[mi]}\int d\,z\,d\,\bar{z}\langle e^{b (e_{k},\varphi(z,\bar{z}))} V_{\alpha_{1}}(z_{1},\bar{z}_{1})...V_{\alpha_{l}}(z_{l},\bar{z}_{l})\rangle
\end{eqnarray}
 Multiplying with $\omega_{o[i]}$, where $\omega^{o}$ are the fundamental weights of the Lie algebra, $(e_{k},\omega_{o})=e_{k[m]}K_{[mi]}\omega_{o[i]}=\delta_{[ko]}$, we get:
\begin{equation}\label{toda1}
(\sum_{j=1}^{l}(\alpha_{j},\omega_{o})-2(Q,\omega_{o}))\langle V_{\alpha_{1}}(z_{1},\bar{z}_{1})...V_{\alpha_{l}}(z_{l},\bar{z}_{l})\rangle=\mu b \int d\,z\,d\,\bar{z}\langle e^{b (e_{o},\varphi(z,\bar{z}))} V_{\alpha_{1}}(z_{1},\bar{z}_{1})...V_{\alpha_{l}}(z_{l},\bar{z}_{l})\rangle
\end{equation}
The dual of this equation is defined with the following replacements: $b\to\frac{1}{b}$ and $\mu\to\tilde{\mu}$ which can be analogously argued using screening charges and thus yielding:
\begin{equation}\label{toda2}
(\sum_{j=1}^{l}(\alpha_{j},\omega_{o})-2(Q,\omega_{o}))\langle V_{\alpha_{1}}(z_{1},\bar{z}_{1})...V_{\alpha_{l}}(z_{l},\bar{z}_{l})\rangle=\tilde{\mu} \frac{1}{b} \int d\,z\,d\,\bar{z}\langle e^{\frac{1}{b}(e_{o},\varphi(z,\bar{z}))} V_{\alpha_{1}}(z_{1},\bar{z}_{1})...V_{\alpha_{l}}(z_{l},\bar{z}_{l})\rangle
\end{equation}
The above two are the screening charge equations in Sl(n) Toda field theory. Using this we will now make some comments about the structure of the three point function in this theory.
\subsection*{Infinite product form}
As earlier we can use the recursion relations to write an infinite product form for the part of the three point function which is the denominator.
If we define $\tilde{c}_{o}=(\sum_{j=1}^{l}(\alpha_{j},\omega_{o})-2(Q,\omega_{o}))$, then the recursion relations yield the following form :
\begin{footnotesize}
\begin{eqnarray}
&& \!\!\!\!\!\! \!\!\!\!\!\! \!\!\!\!\!\!\!\!\!\!\!\! \!\!\!\!\!\!\!\!\!\!\!\!
\lim_{k_{i},p_{i}\to\infty}\prod_{m_{1}=0}^{k_{1}}\dots\prod_{m_{n-1}=0}^{k_{n-1}}\prod_{l_{1}=0}^{p_{1}}\dots\prod_{l_{n-1}=0}^{p_{n-1}}(\tilde{c}_{1}+bm_{1}+\frac{1}{b}l_{1})\dots(\tilde{c}_{n-1}+bm_{n-1}+\frac{1}{b}l_{n-1})\langle V_{\alpha_{1}}(z_{1},\bar{z}_{1})...V_{\alpha_{l}}(z_{l},\bar{z}_{l})\rangle\notag\\
&& \!\!\!\!\!\! \!\!\!\!\!\! \!\!\!\!\!\!\!\!\!\!\!\! \!\!\!\!\!\!\!\!\!\!\!\!
=\lim_{k_{i},p_{i}\to\infty}(\mu b)^{k_{1}+\dots+k_{n-1}}(\frac{\tilde{\mu}}{b})^{p_{1}+\dots+p_{n-1}}\prod_{m_{1}=0}^{k_{1}}\dots\prod_{m_{n-1}=0}^{k_{n-1}}\prod_{l_{1}=0}^{p_{1}}\dots\prod_{l_{n-1}=0}^{p_{n-1}} \int d\,z_{m_{1}}\,d\,\bar{z}_{m_{1}}\dots d\,z_{m_{n-1}}\,d\,\bar{z}_{m_{n-1}}d\,z_{l_{1}}\,d\,\bar{z}_{l_{1}}\dots\notag\\
&& \!\!\!\!\!\! \!\!\!\!\!\! \!\!\!\!\!\!\!\!\!\!\!\! \!\!\!\!\!\!\!\!\!\!\!\!
 d\,z_{l_{n-1}}\,d\,\bar{z}_{l_{n-1}}\langle e^{b(e_{1},\varphi(z_{m_{1}},\bar{z}_{m_{1}}))}\dots e^{b(e_{n-1},\varphi(z_{m_{n-1}},\bar{z}_{m_{n-1}}))} e^{\frac{1}{b}(e_{1},\varphi(z_{l_{1}},\bar{z}_{l_{1}}))}\dots e^{\frac{1}{b}(e_{n-1},\varphi(z_{l_{n-1}},\bar{z}_{l_{n-1}}))} V_{\alpha_{1}}(z_{1},\bar{z}_{1})...V_{\alpha_{l}}(z_{l},\bar{z}_{l})\rangle\notag\\
\end{eqnarray}
\end{footnotesize}
This is due to the orthogonality relation $(e_{k},\omega_{o})=e_{k[m]}K_{[mi]}\omega_{o[i]}=\delta_{[ko]}$. Hence shift in any $\tilde{c}_{k}$'s by $b\,e_{k}$ or $\frac{1}{b}\,e_{k}$ is of the type: $\tilde{c}_{k}+b(e_{k},\omega_{k})$, e.t.c.. Thus in the limit of large iterations one obtains the infinite product for the first part of the function:
\begin{equation}
\prod_{m_{1}=0}^{\infty}\dots\prod_{m_{n-1}=0}^{\infty}\prod_{l_{1}=0}^{\infty}\dots\prod_{l_{n-1}=0}^{\infty}(\tilde{c}_{1}+bm_{1}+\frac{1}{b}l_{1})\dots(\tilde{c}_{n-1}+bm_{n-1}+\frac{1}{b}l_{n-1})
\end{equation}

This infinite product can be seen to be coming from product of the inverses of the Barnes Double Gamma functions: $\prod_{k=0}^{n-1}\Gamma(\tilde{c}_{k};b,\frac{1}{b})$ again by using Weistrass theorem. In analogy to Liouville theory, one has to construct the duals to this products which are connected by the Weyl group of Sl(n). The action of the Weyl group in the case of Toda field thoeries is given as follows \cite{fateev1}:
\begin{equation}\label{weyl}
s(\alpha)=Q+\hat{s}(\alpha-Q)
\end{equation}
where $\hat{s}$ is a member of the Weyl group of Sl(n). For simple reflection we obtain $s(\alpha-2Q)=4Q-\alpha$. Using the analogy to Liouville theory we proposes that the denominator is a Weyl invariant entire function \cite{fateev1},\cite{Fateev}. For the case of simple reflection we must have another product of double Gamma functions : $\prod_{k=0}^{n-1}\Gamma(\tilde{c}^{*}_{k};b,\frac{1}{b})$, where $\tilde{c}^{*}_{k}=((4Q-\tilde{\alpha}),\omega_{k})$. Hence one would expect that the full form would compose of products of Barnes double gamma functions with arguments which are connected by action of the Weyl group, (namely $(s(\tilde{\alpha}-2Q),\omega_{k})$, e.t.c., with different $\hat{s}$). As the Weyl group is finite, hence the actions  would terminate after a finite number. Thus the Weyl group basically permutes the arguments of the double gamma functions, leaving the product invariant. Unfortunately it is technically difficult to guess for a large group Sl(n), although we intend to look at this problem in our future work. One can directly see from the equation (\ref{toda1}) or (\ref{toda2}), by action of the Weyl group on one of the $\alpha_{i}$ that:
\begin{eqnarray}
&& \!\!\!\!\!\! \!\!\!\!\!\! \!\!\!\!\!\!\!\!\!\!\!\! \!\!\!\!\!\!\!\!\!\!\!\!
(\sum_{j=1,j\neq i}^{l}(\alpha_{j}+s_{k}(\alpha_{i}),\omega_{o})-2(Q,\omega_{o}))\langle V_{\alpha_{1}}(z_{1},\bar{z}_{1})..R_{s_{k}}^{-1}(\alpha_{i})V_{\alpha_{i}}(z_{i}\bar{z}_{i})..V_{\alpha_{l}}(z_{l},\bar{z}_{l})\rangle\notag\\
&& \!\!\!\!\!\! \!\!\!\!\!\! \!\!\!\!\!\!\!\!\!\!\!\! \!\!\!\!\!\!\!\!\!\!\!\!
\quad\quad\quad\quad\quad\quad=\mu b \int d\,z\,d\,\bar{z}\langle e^{b (e_{o},\varphi(z,\bar{z}))} V_{\alpha_{1}}(z_{1},\bar{z}_{1})..R_{s_{k}}^{-1}(\alpha_{i})V_{\alpha_{i}}(z_{i}\bar{z}_{i})..V_{\alpha_{l}}(z_{l},\bar{z}_{l})\rangle
\end{eqnarray}
Where the reflection symmetry is used, \cite{Fateev}:
\begin{equation}
V_{\alpha}(z,\bar{z})=R_{s}(\alpha)V_{s(\alpha)}(z,\bar{z})
\end{equation}
One can cancel the reflection amplitudes from both sides yielding:
\begin{eqnarray}\label{weyl1}
&& \!\!\!\!\!\! \!\!\!\!\!\! \!\!\!\!\!\!\!\!\!\!\!\! \!\!\!\!\!\!\!\!\!\!\!\!
(\sum_{j=1,j\neq i}^{l}(\alpha_{j}+s_{k}(\alpha_{i}),\omega_{o})-2(Q,\omega_{o}))\langle V_{\alpha_{1}}(z_{1},\bar{z}_{1})..V_{\alpha_{i}}(z_{i}\bar{z}_{i})..V_{\alpha_{l}}(z_{l},\bar{z}_{l})\rangle\notag\\
&& \!\!\!\!\!\! \!\!\!\!\!\! \!\!\!\!\!\!\!\!\!\!\!\! \!\!\!\!\!\!\!\!\!\!\!\!
\quad\quad\quad\quad\quad\quad=\mu b \int d\,z\,d\,\bar{z}\langle e^{b (e_{o},\varphi(z,\bar{z}))} V_{\alpha_{1}}(z_{1},\bar{z}_{1})..V_{\alpha_{i}}(z_{i}\bar{z}_{i})..V_{\alpha_{l}}(z_{l},\bar{z}_{l})\rangle
\end{eqnarray}
One can do this again with a different $s$ namely $s_{m}$, and hence following similar arguments we arrive:
\begin{eqnarray}\label{weyl2}
&& \!\!\!\!\!\! \!\!\!\!\!\! \!\!\!\!\!\!\!\!\!\!\!\! \!\!\!\!\!\!\!\!\!\!\!\!
(\sum_{j=1,j\neq i}^{l}(\alpha_{j}+s_{m}s_{k}(\alpha_{i}),\omega_{o})-2(Q,\omega_{o}))\langle V_{\alpha_{1}}(z_{1},\bar{z}_{1})..V_{\alpha_{i}}(z_{i}\bar{z}_{i})..V_{\alpha_{l}}(z_{l},\bar{z}_{l})\rangle\notag\\
&& \!\!\!\!\!\! \!\!\!\!\!\! \!\!\!\!\!\!\!\!\!\!\!\! \!\!\!\!\!\!\!\!\!\!\!\!
\quad\quad\quad\quad\quad\quad=\mu b \int d\,z\,d\,\bar{z}\langle e^{b (e_{o},\varphi(z,\bar{z}))} V_{\alpha_{1}}(z_{1},\bar{z}_{1})..V_{\alpha_{i}}(z_{i}\bar{z}_{i})..V_{\alpha_{l}}(z_{l},\bar{z}_{l})\rangle
\end{eqnarray}
Equation (\ref{weyl1}) is exactly similar to (\ref{toda1}) with replacement of $\alpha_{i}$ by $s_{k}(\alpha_{i})$ in the first part of the L.H.S. , $k$ denoting one of the reflection elements of the Weyl group. Similarly for (\ref{weyl2}), since the elements $s_{k},\,s_{m}$ are elements of the Weyl group this is equivalent to some $s_{l}$ under group multiplication. 
 Thus in general one can write this equation for all possible Weyl group elements. So one has,
\begin{eqnarray}
&& \!\!\!\!\!\! \!\!\!\!\!\! \!\!\!\!\!\!\!\!\!\!\!\! \!\!\!\!\!\!\!\!\!\!\!\!
(\sum_{j=1,j\neq i}^{l}(\alpha_{j}+ s_{k}(\alpha_{i}),\omega_{o})-2(Q,\omega_{o}))\langle V_{\alpha_{1}}(z_{1},\bar{z}_{1})..V_{\alpha_{i}}(z_{i}\bar{z}_{i})..V_{\alpha_{l}}(z_{l},\bar{z}_{l})\rangle\notag\\
&& \!\!\!\!\!\! \!\!\!\!\!\! \!\!\!\!\!\!\!\!\!\!\!\! \!\!\!\!\!\!\!\!\!\!\!\!
\quad\quad\quad\quad\quad\quad=\mu b \int d\,z\,d\,\bar{z}\langle e^{b (e_{o},\varphi(z,\bar{z}))} V_{\alpha_{1}}(z_{1},\bar{z}_{1})..V_{\alpha_{i}}(z_{i}\bar{z}_{i})..V_{\alpha_{l}}(z_{l},\bar{z}_{l})\rangle\quad, s_{k}\in \it{W}
\end{eqnarray}
Where $W$ means Weyl group.
 Similar equation exists for $\frac{1}{b}$ and using these, analogous infinite products can be obtained, suggesting the duality of the arguments of the double gamma functions via Weyl symmetry under assumption that the denominator is an entire function and with zeroes in the variables $\tilde{c}_{o}=mb+n/b$ only, $m$ and $n$ being non negative integers. From these arguments we suggest the following form for the denominator of the three point function:
\begin{equation}
\prod_{s_{k}s_{l}s_{m}\in \it{W}}\frac{1}{\prod_{o=1}^{n-1}\Gamma_{2}\bigg( s_{k}(\alpha_{1})+ s_{l}(\alpha_{2})+ s_{m}(\alpha_{3})-2Q),\omega_{o}\bigg)}
\end{equation}
Where $s_{k},s_{l},s_{m}$ are the Weyl group elements for Sl(n)  i.e. $\in \it{W}$. It must be stated here that there might exist some remaining part of the denominator which is dual in $b$ and $1/b$ and also Weyl invariant which can not be deduced here, as the analysis explicitly makes use of the pole structure to obtain the above expression. Nevertheless as discussed in \cite{fateev2}, the denominator is a symmetric and Weyl invariant function of the variables $\alpha_{1},\,\alpha_{2},\,\alpha_{3}$, and thus as one of the factors derived from the infinite product form is $\prod_{k=0}^{n-1}\Gamma(\tilde{c}_{k};b,\frac{1}{b})$, ($\tilde{c}_{k}$ being defined for three $\alpha$s) for total Weyl invariance, the other Barnes Double gamma functions must also be there in the denominator. 

One must also note a key feature of the structure of the three point function. The three point function in conformal field theory is a symmetric function of the parameters $\alpha$s. This is because the three points on the sphere $0,1,\infty$ are interchangable by conformal transformation. If one looks at Liouville theory, then one can note that the denominator is a completely Weyl invariant function, as the expression,
\begin{equation}
\Upsilon(\alpha_{1}+\alpha_{2}+\alpha_{3}-Q)\Upsilon(\alpha_{1}+\alpha_{2}-\alpha_{3})\Upsilon(\alpha_{2}+\alpha_{3}-\alpha_{1})\Upsilon(\alpha_{3}+\alpha_{1}-\alpha_{2})
\end{equation}
is in principle equal to products of inverse Barnes double gamma functions. It is easy to see from them that each of the double gamma functions are connected by Weyl reflections in each of  the variables $\alpha_{i}$, $(\alpha_{i}\to Q-\alpha_{i})$ (i.e. permutation of the argument of the double gamma functions and hence simple algebraic permutations, the function being the same). In the case of Toda field theory, the simple reflection is substituted by action of all possible Weyl group element which was discussed in the previous paragraph.

\section*{Sine Liouville Theory}
Now we move onto examine another model namely Sine-Liouvile thoery, using the zero mode technique discussed. The action for conformally invariant Sine-Liouville theory is given as follows \cite{Rim}:
\begin{equation}
\frac{1}{16\pi}\int d^{2}x \bigg\{(\partial_{\mu}\phi_{1})^{2}+(\partial_{\mu}\phi)_{2})^{2}+(\partial_{\mu}\phi_{3})^{2}-32\mu e^{\alpha\phi_{1}}cos(\beta\phi_{2}+\gamma\phi_{3})+q\phi_{1}R\bigg\}
\end{equation}
Where $q=\frac{1}{2\alpha}(1+\alpha^{2}-\beta^{2}-\gamma^{2})$ is the background charge 
 and $R$ is the two-dimensional scalar curvature. A normal ordered vertex operator at point $z,\bar{z}$ is denoted as:
\begin{equation}
V(\vec{a};z)= :exp[a\phi_{1}+i(b\phi_{2}+c\phi_{3})]:(z,\bar{z})
\end{equation}
A string of these vertex operators can be understood as delta function kind of sources:
\begin{equation}
J_{\phi_{1}}(x)=\sum_{i=1}^{n}a_{i}\delta^{2}(x-x_{i})\quad J_{\phi_{2}}(x)=\sum_{i=1}^{n}ib_{i}\delta^{2}(x-x_{i})\quad  J_{\phi_{3}}(x)=\sum_{i=1}^{n}ic_{i}\delta^{2}(x-x_{i})
\end{equation}

Now translation invariance with respect to the zero mode of $\phi_{1}$ yields the following equation:
\begin{equation}
(\sum_{i=1}^{n}a_{i}-2q)\langle\prod_{i=1}^{n}V(\vec{a}_{i};z_{i})\rangle=2\mu\alpha\int d^{2}x\langle e^{\alpha\phi_{1}(x)}cos(\beta\phi_{2}+\gamma\phi_{3})V(\vec{a}_{i};z_{i})\rangle
\end{equation}
Note this equation is very similar to what one obtains in Liouville Field theory.
Next let us perform the same with the zero mode of $\phi_{2}$, $\phi_{3}$ :
\begin{eqnarray}
&&(\sum_{i=1}b_{i})\langle\prod_{i=1}^{n}V(\vec{a}_{i};z_{i})\rangle=\\
&&\frac{1}{2}\int d^{2}x\,\bigg[ 2\mu\beta\langle e^{\alpha\phi_{1}(x)+i(\beta\phi_{2}(x)+\gamma\phi_{3}(x))}\prod_{i=1}^{n}V(\vec{a}_{i};z_{i})\rangle-2\mu\beta\langle e^{\alpha\phi_{1}(x)-i(\beta\phi_{2}(x)+\gamma\phi_{3}(x))}\prod_{i=1}^{n}V(\vec{a}_{i};z_{i})\rangle\bigg]\notag
\end{eqnarray}
And 
\begin{eqnarray}
&&(\sum_{i=1}c_{i})\langle\prod_{i=1}^{n}V(\vec{a}_{i};z_{i})\rangle=\\
&&\frac{1}{2}\int d^{2}x\,\bigg[ 2\mu\gamma\langle e^{\alpha\phi_{1}(x)+i(\beta\phi_{2}(x)+\gamma\phi_{3}(x))}\prod_{i=1}^{n}V(\vec{a}_{i};z_{i})\rangle-2\mu\gamma\langle e^{\alpha\phi_{1}(x)-i(\beta\phi_{2}(x)+\gamma\phi_{3}(x))}\prod_{i=1}^{n}V(\vec{a}_{i};z_{i})\rangle\bigg]\notag
\end{eqnarray}
Using these two one can obtain the following condition:
\begin{equation}
(\frac{1}{\beta}\sum_{i=1}^{n}b_{i}-\frac{1}{\gamma}\sum_{i=1}^{n}c_{i})\langle\prod_{i=1}^{n}V(\vec{a}_{i};z_{i})\rangle=0
\end{equation}
since, the $U(1)$ charge of a vertex operator is given by $Q=b/\beta-c/\gamma$, this is a $U(1)$ charge neutrality constraint i.e.
\begin{equation}
\sum_{i=1}^{n}Q_{i}=0
\end{equation}
Again after doing some more algebra we obtain:
\begin{eqnarray}
&&\frac{(\beta\sum_{i}^{n}b_{i}+\gamma\sum_{i=1}^{n}c_{i})}{\beta^{2}+\gamma^{2}}\langle\prod_{i=1}^{n}V(\vec{a}_{i};z_{i})\rangle=\\
&&\frac{1}{2}\int d^{2}x\, \bigg[2\mu\langle e^{\alpha\phi_{1}(x)+i(\beta\phi_{2}(x)+\gamma\phi_{3}(x))}\prod_{i=1}^{n}V(\vec{a}_{i};z_{i})\rangle-2\mu\langle e^{\alpha\phi_{1}(x)-i(\beta\phi_{2}(x)+\gamma\phi_{3}(x))}\prod_{i=1}^{n}V(\vec{a}_{i};z_{i})\rangle\bigg]\notag
\end{eqnarray}
The winding number of a vertex operator is defined as $\omega=(\beta b+\gamma c)/(\beta^{2}+\gamma^{2})$, and so the aove equation can be written as folows:
\begin{eqnarray}
&&(\sum_{i=1}^{n}\omega_{i})\langle\prod_{i=1}^{n}V(\vec{a}_{i};z_{i})\rangle=\\
&&\frac{1}{2}\int d^{2}x\,\bigg[ 2\mu\langle e^{\alpha\phi_{1}(x)+i(\beta\phi_{2}(x)+\gamma\phi_{3}(x))}\prod_{i=1}^{n}V(\vec{a}_{i};z_{i})\rangle-2\mu\langle e^{\alpha\phi_{1}(x)-i(\beta\phi_{2}(x)+\gamma\phi_{3}(x))}\prod_{i=1}^{n}V(\vec{a}_{i};z_{i})\rangle\bigg]\notag
\end{eqnarray}
Let us define the following
\begin{equation}
\sum_{i=1}^{n}a_{i}=\tilde{a}\quad,\quad\sum_{i=1}^{n}Q_{i}=\tilde{Q}\quad,\quad\sum_{i=1}^{n}\omega_{i}=\tilde{\omega}
\end{equation}
One can do some more algebra and then arrive at the following equations for the correlation functions:
\begin{eqnarray}
&&(\frac{(\tilde{a}-2q)}{\alpha}+\tilde{\omega})\langle\prod_{i=1}^{n}V(\vec{a}_{i};z_{i})\rangle=2\mu\int d^{2}x\,\langle e^{\alpha\phi_{1}(x)+i(\beta\phi_{2}(x)+\gamma\phi_{3}(x))}\prod_{i=1}^{n}V(\vec{a}_{i};z_{i})\rangle\\
&&(\frac{(\tilde{a}-2q)}{\alpha}-\tilde{\omega})\langle\prod_{i=1}^{n}V(\vec{a}_{i};z_{i})\rangle=2\mu\int d^{2}x\,\langle e^{\alpha\phi_{1}(x)-i(\beta\phi_{2}(x)+\gamma\phi_{3}(x))}\prod_{i=1}^{n}V(\vec{a}_{i};z_{i})\rangle\\
&&\tilde{Q}\langle\prod_{i=1}^{n}V(a_{i};z_{i})\rangle=0
\end{eqnarray}
These three equation are the screening charge and charge neutrality equations. Of course in analogy to Liouville theory there exists more screening charges. The conformal dimension of a vertex operator is given $\Delta=a(2q-a)+b^{2}+c^{2}$. Hence one can write down the dual screening charge equations : 
\begin{eqnarray}
&&(\frac{(\tilde{a}-2q)}{2q-\alpha}+\tilde{\omega})\langle\prod_{i=1}^{n}V(\vec{a}_{i};z_{i})\rangle=2\tilde{\mu}\int d^{2}x\,\langle e^{(2q-\alpha)\phi_{1}(x)+i(\beta\phi_{2}(x)+\gamma\phi_{3}(x))}\prod_{i=1}^{n}V(\vec{a}_{i};z_{i})\rangle\\
&&(\frac{(\tilde{a}-2q)}{2q-\alpha}-\tilde{\omega})\langle\prod_{i=1}^{n}V(\vec{a}_{i};z_{i})\rangle=2\tilde{\mu}\int d^{2}x\,\langle e^{(2q-\alpha)\phi_{1}(x)-i(\beta\phi_{2}(x)+\gamma\phi_{3}(x))}\prod_{i=1}^{n}V(\vec{a}_{i};z_{i})\rangle
\end{eqnarray}
Where $\tilde{\mu}$ is some dual cosmological constant. In analogy to Liouville theory, one has to look for a particular function which satisfies these functional equations. 
Let us look at one of the equations, after itterations,
\begin{eqnarray}
&&\prod_{j=0}^{m}(\frac{(\tilde{a}-2q)}{\alpha}+\tilde{\omega}+2j)\langle\prod_{i=0}^{n}V(\vec{a}_{i};z_{i})\rangle\notag\\
&&\quad\quad\quad\quad\quad\quad=(2\mu)^{m+1}\prod_{j=0}^{m}\int d^{2}x_{j}\,\langle e^{\alpha\phi_{1}(x_{j})+i(\beta\phi_{2}(x_{j})+\gamma\phi_{3}(x_{j}))}\prod_{i=1}^{n}V(\vec{a}_{i};z_{i})\rangle\notag
\end{eqnarray}
This is similar to what we had for Liouville field theory in \cite{pd}. To solve these functional equations, one has to look for a particular function which has these properties (just like the Barnes double gamma function solves the functional equations in the case of Liouville theory)

\subsection*{Infinite product expansion}
Here we write down the infinite product form that we obtain by iterative scheme for the functional equation obtained above for Sine-Liouville theory. The form is as follows:
\begin{equation}
\prod_{j=0}^{\infty}\prod_{m=0}^{\infty}((\tilde{a}-2q)+(\tilde{\omega}+2j)\alpha+(\tilde{\omega}+2m)(2q-\alpha))\prod_{l=0}^{\infty}\prod_{n=0}^{\infty}((\tilde{a}-2q)+(-\tilde{\omega}+2l)\alpha+(-\tilde{\omega}+2n)(2q-\alpha))
\end{equation}

One has to look for an entire function which has the above infinite product expansion to deduce the denominator of the three point function in particular. The information about the rest of the piece, must be obtained from the residues as usual and also from Weyl reflection of the Liouville part i.e. in the variable $a$s.

\section*{Conclusion}
In this article we showed that if one starts with equation of motion in presence of all the screening charges in a conformal field theory (2 in case of Liouville), then there exists set of dual functional zero mode Schwinger Dyson equations for all the screening charges. Also Using the path integral for the Liouville theory derived using the operator equation of motion for two screening charges, we obtained the familiar pole structure for the correlation functions. Using the analogy to Liouville theory, we investigated the correlation functions first in the case of $Sl(n)$ Toda field theories and then Sine-Liouville theory using the zero mode functional equations to obtain information about their structure which was the primary motivation of this work. In the first case a general structure for the denominator composed of entire functions was obtained for the Toda field theory case. In the second case an infinite product form was obtained but the form of the entire function could not be deduced. In future we would like to build upon this work, in order to construct a possible algebraic technique to compute structure constants in some specific conformal field theory models.

\section*{Acknowledgement}
The author would like to thank Stefan Fredenhagen for useful comments and suggestions about the draft. The author would also like to thank George Jorjadze and Suvankar Dutta for discussions and comments.

\end{document}